\documentstyle[editedbook,epsfig,psfig,epsf]{mq}

\def\msun{$M_{\odot}$ }
\def\mdot{$\dot{M}$ }
\def\msunyr{M$_{\odot}$ yr$^{-1}$ }

\def\ergs{erg s$^{-1}$ }

\def\cm2{cm$^2$ }
\def\se1{s$^{-1}$ }

\begin{opening}
\title{The Duty Cycle of GRS 1915+105}
\author{O. Vilhu \& D. Hannikainen}
\institute{ Observatory, University of Helsinki, Finland}
\end{opening}

\runningtitle{Duty cycle of GRS 1915+105}
\runningauthor{Vilhu \& Hannikainen}

\begin{document}
\vspace{-0.5cm}
\begin{abstract}
{\small We propose a scenario for a periodic filling and emptying of the accretion disc
of GRS 1915+105, by computing the mass transfer rate from the donor and comparing it
with the observed accretion rate (see the full paper \cite{Vil01}). The binary parameters found
by  \cite{Gre01} predict evolutionary expansion of the donor along the 
giant branch
with a conservative mass transfer rate (1 -- 2) $\times$ 10$^{-8}$ \msunyr. This reservoir
can support the present observed accretion rate with a duty cycle 0.05 -- 0.1 (the active
time as a fraction of the total life time). The viscosity time scale at the 
circularization radius (15 solar radii from the primary 14 \msun black hole)
 is identified as the recurrent quiescent time during which a new disc is formed 
once consumed by the  BH. For small
viscosity ($\alpha$ = 0.001) it equals to 300 -- 400 years. 
The microquasar phase,
with the duty cycle, will last around 10$^7$ years 
ending with a long-period black hole +
white dwarf system. 
     }
\end{abstract}

\section{Results}

Greiner et al. \cite{Gre01}   identified a K-M III giant as  
the mass-donating component of
GRS 1915+105  in a 33.5 day orbit around a 14 \msun black
hole (see Fig. 1). We assume that the donor fills its Roche lobe and that the mass loss is
determined by evolutionary expansion along the giant branch, conserving total
mass and angular momentum. The growth of the He-core, resulting in an increase of the radius,
is determined by the luminosity due to hydrogen shell burning which depends completely
on the core mass \cite{Web83}.

Assuming the donor mass is around 1.2 \msun it is rather easy to
estimate the mass transfer from the donor  (1 -- 2) $\times$ 10$^{-8}$ \msunyr
(Fig. 2 and Table 1, for details see \cite{Vil01}). 

The mean X-ray luminosity over the past 10 years (1.2$\times$10$^{39}$ \ergs using
a distance of 12.5 kpc, see
Fig. 3) gives an accretion rate of (0.2/$\eta$) 10$^{-7}$ \msunyr onto
the primary black hole where $\eta$ is the conversion factor of mass infall into radiation
(L = $\eta$ \mdot c$^2$).  Finally, the ratio (mass transfer/accretion) gives the
duty cycle {\bf (0.05 - 0.1)($\eta$/0.1)}, the active time as a fraction 
of the total life time. 

We suggest that the quiescent OFF-time is the viscosity time scale at the
circularization radius (15 solar radii from the BH) where the local
Keplerian angular momentum equals to that in the L1-point. For small
$\alpha$ (= 0.001) the time scale equals to 300 -- 400 years. In this particular case
the present active phase may last 10 -- 20 years more.

The whole microquasar phase may have started when the  secondary made its
first contact with its Roche lobe just after finishing the main sequence phase
or while crossing the Hertzsprung gap, in a 10-day binary. The final 
end product  after  10$^7$ years
will probably be a long-period detached
white dwarf -- black hole system.

\begin{figure}[htb]
\centering
\psfig{file=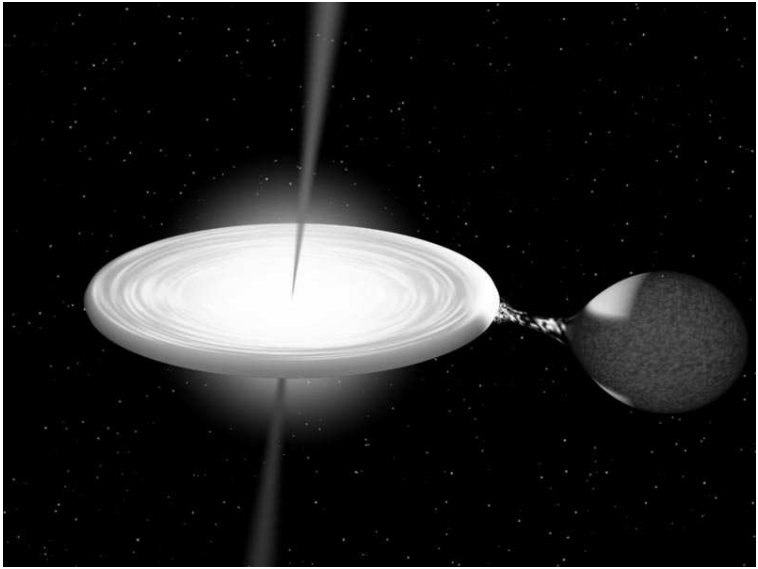,width=10cm}
\caption{BinSim view of GRS 1915+105 (courtesy of Panu Muhli). BinSim was 
created by Rob Hynes (www.astro.soton.ac.uk). The low-mass K-giant donor fills its 
Roche lobe in a 33.5 day orbit around the 14 \msun primary black hole.
The disc extends up to four times the circularization radius (twice angular momentum there)
and may represent maximum filling of the disc and the start of the duty cycle.}
\label{fig:binsim}
\end{figure}


\begin{figure}[htb]
\centering
\psfig{file=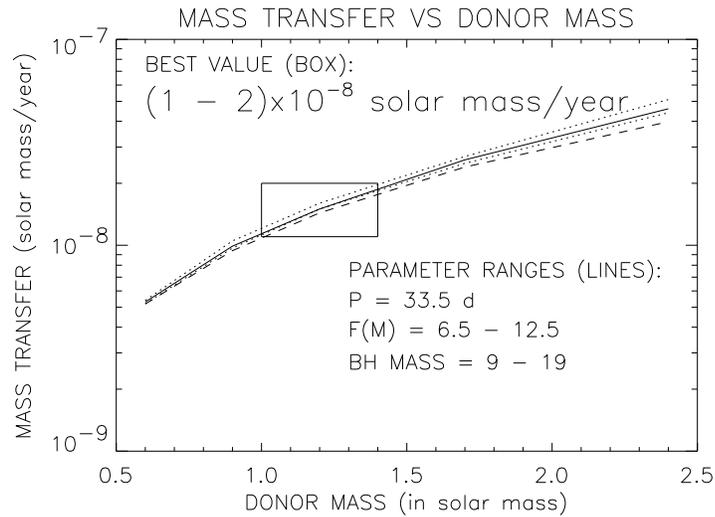,width=10cm}
\caption{Computed mass transfer rate from the donor depending on its assumed mass. The box shows the mass range suggested by Greiner et al. } 
                       
\label{fig:transfer}
\end{figure}

\begin{table}[htbp]
\centering
\begin{tabular}{|c c c c c c c c|}
\hline
\multicolumn{8}{|c|}{Donor parameters}\\
\hline
Sp & M$_K$ & M$_{He}$ & L & R & \mdot & a & R$_{circ}$  \\
\hline
K6 & -2.6 & 0.28 & 77 & 21 & 1.5 10$^{-8}$ & 108 & 14 \\
K5-M1 & -2.2 - -2.7 & 0.26 - 0.29 & 50 - 100 & 17 - 27 & 5 10$^{-9}$ -
5 10$^{-8}$ & 95 - 115 & 12 - 18 \\
\hline
\end{tabular}

\caption{ The first line gives the computed parameters for a 1.2 \msun donor
(in solar units,  \mdot is given in \msunyr).  
The second line gives the ranges if the donor mass is varied 
between 0.6 - 2.4 \msun, within the mass function limit. Sp = spectral
type, M$_K$ = absolute K-magnitude, M$_{He}$ = mass of the He-core, 
L = luminosity, R = radius, \mdot = mass transfer from the donor,
a = binary separation, R$_{circ}$ = circularization radius.  }
\label{table:donor}
\end{table}
\begin{figure}[htb]
\centering
\psfig{file=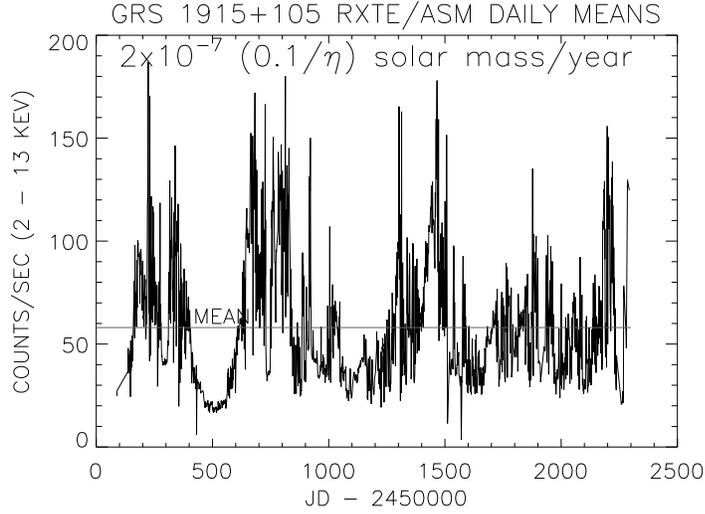,width=10cm}
\caption{ASM light curve of GRS 1915+105. 
The text gives the mean accretion rate onto the primary BH where 
$\eta$ is the conversion of mass infall into radiation. }
\label{fig:accr}
\end{figure}

\section*{ Acknowledgments}.

This research is part of the consortium `High Energy Astrophysics
and Space Astronomy' financed by the Academy of Finland and the Finnish
Technology Agency TEKES within the framework of the Space
Research program ANTARES. \\


\begin{thebibliography}{}

\bibitem{Gre01}
 Greiner J., Cuby I.G. \&  McCaughrean M.J., 2001, astro-ph/0111538.
\bibitem{Web83}
Webbink, R.F., Rappaport, S. \& Savonije G.J. 1983, ApJ {\bf 270}, 678.
\bibitem{Vil01}
Vilhu, O. 2001, Astronomy and Astrophysics (in press), astro-ph/0204146\\

\end{thebibliography}
\end{document}